\documentclass[preprint,showpacs,preprintnumbers,amsmath,amssymb]{revtex4}

\usepackage{graphicx}
\usepackage{dcolumn}
\usepackage{bm}
\usepackage{epstopdf}

\begin{document}

\title{The role of internal chain dynamics on the rupture kinetics\\
of adhesive contacts}

\author{V. Barsegov$^1$, G. Morrison$^{2}$ and D. Thirumalai$^{2}$}
\affiliation{$^1$Department of Chemistry, University of Massachusetts, Lowell, MA 01854\\
$^2$Biophysics Program, Institute for Physical Science and Technology, 
University of Maryland, College Park, MD 20742}


\begin{abstract}

\noindent 
We study the forced rupture of adhesive contacts between monomers that are not 
covalently linked in a Rouse chain. When the applied force ($f$) to the chain end 
is less than the critical force for rupture ($f_c$), the {\it reversible} rupture
process is coupled to the internal Rouse modes. If  $f/f_{c}$$>$$1$ the rupture 
is {\it irreversible}. In both limits, the non-exponential distribution of contact 
lifetimes, which depends sensitively on the location of the contact, follows the 
double-exponential (Gumbel) distribution. When two contacts are well separated 
along the chain, the rate limiting step in the {\it sequential} rupture kinetics is the 
disruption of the contact that is in the chain interior. If the two contacts are 
close to each other, they cooperate to sustain the stress, which results in an
``all-or-none'' transition.  

\end{abstract}
\maketitle


Nanomanipulation of biological  molecules by force can be used to  explore and control 
their energy landscapes at the single molecule level
\cite{BlockAnnRevBiophys2007TinocoPNAS07FernandezPNAS07}. 
The challenge is to solve the inverse problem, namely, to extract the unbinding pathways 
and structural features pertaining to the systems of interest from measurements such as the 
force-extension curves, lifetime distribution of adhesive contacts, and unbinding force distributions. 
From these data, many characteristics of the energy landscape (roughness, barrier height, 
and movement of the transition states) have been obtained 
using  theoretical models \cite{HyeonPNAS03DudkoPNAS03,BarsegovPRL05}.  
In these models, the force-induced unbinding transitions from the bound state (for 
example folded conformations of RNA) are presumed to occur \textit{irreverisbly} along a 
one dimensional reaction coordinate, $R$, that is conjugate to the applied force, $f$. In this 
picture, $f$ ``tilts" the energy landscape, thus rendering the bound state unstable when $f$ 
exceeds a critical value. On the other hand, detailed simulations of forced unfolding of proteins 
and RNA \cite{Hyeon06BiophysJ} show that collective topology-dependent events that 
describe the internal dynamics \cite{BarsegovPRL05} are  coupled to $R$. Thus, it is important 
to develop solvable models for force-induced disruption of ``folded" structures starting 
from a well-defined Hamiltonian.   

In this letter, we illustrate the interplay between internal polymer  dynamics and the rupture 
kinetics of intramolecular adhesive contacts between monomers.   Such a model is a caricature of 
forced-unfolding of RNA hairpins \cite{BlockAnnRevBiophys2007TinocoPNAS07FernandezPNAS07} 
where the adhesive contact mimics the stability of the stem.
Using an exactly solvable Rouse model, with one or two adhesive contacts between monomers 
that are not covalently linked,  we address the following questions: (i) For a single adhesive 
contact, how do the lifetime distribution functions vary with the backbone distance between 
monomers $r$ and $s$ that are in adhesive contact? (ii) To what extent is the binding of a 
ruptured contact influenced by $f$ and the internal polymer motions? 


{\it The model:}
Consider a Rouse chain with $N$ identical monomers, connected by a harmonic spring with 
the equilibrium distance $a$, with a single adhesive contact $(r,s)$ between 
two non-covalently linked monomers $r$ and $s$, $1$$\le$$r$$<$$s$$\le$$N$. The first monomer 
is fixed, and a constant force ${\bf f}$$=$$f$${\bf x}$ is applied to the $N^{th}$ monomer in the 
direction parallel to the end-to-end vector ${\bf X}$. Letting ${\bf R}_m$$=$$\{ R_m^x,R_m^y,R_m^z \}$ 
be the monomer position, ($m$$=$$1,2,\ldots,N$),  
the equation of  motion is,
$\xi {{d}\over {dt}}{\bf R}_m$$=$$-\nabla_{{\bf R}_m}$$H(\{ {\bf R}_m \})$$+$${\bf F}_m$
where $\nabla_{{\bf R}_m}$$\equiv$$\partial/\partial {\bf R}_m$, $\xi$ is the friction cefficient, 
and ${\bf F}_m$ is the random force with zero mean and 
$\langle F_m^{\alpha}(t)F_n^{\beta}(0)\rangle $$=$$2\xi k_B T\delta_{mn}\delta_{\alpha \beta}\delta(t)$
($\alpha$, $\beta$$=$$x$, $y$, $z$). We model the  adhesive contact $(r,s)$ 
using a harmonic potential with a spring constant, $\kappa$$=$$3k_B T/b^2$, where $b$ is the 
equilibrium contact distance. The Hamiltonian with a single contact under tension is taken to 
be $H$$=$$H_0$$+$${\bf f_{N}}({\bf R}_N-{\bf R}_1)$ where
\begin{equation}\label{1.3}
H_0 = {{3k_B T}\over {2a^2}}\sum_{n=2}^{N}({\bf R}_n-{\bf R}_{n-1})^2 +  {{3k_B T}\over {2b^2}}\Theta(B-|{\bf R}_{rs}|){\bf R}_{rs}^2
\end{equation}
with ${\bf R}_{rs}$$=$${\bf R}_{r}$$-$${\bf R}_{s}$, and $\Theta(x)$ is the Heaviside step function. The 
contact is ruptured when $|{\bf R}_{rs}|$$\ge$$B$.  The probability density function (pdf) of the 
$(r,s)$-contact lifetimes is obtained as  $p_{rs}(t,f)$$=$$-d P_{rs}(t,f)/dt$,
where $P_{rs}(t)$,  the probability that the adhesive contact remains intact, is 
$P_{rs}(t,f)$$=$${{\int_{|{\bf R_{rs}}|\le B} d^3{\bf R}_{rs} 
\int d^3{\bf R}_{rs}^0 Q_{f}({\bf R}_{rs},t;{\bf R}_{rs}^0) P_{I}({\bf R}_{rs}^0)}
\over {\int d^3{\bf R}_{rs}
\int d^3{\bf R}_{rs}^0 Q_{f}({\bf R}_{rs},t;{\bf R}_{rs}^0)P_{I}({\bf R}_{rs}^0)}}$.
Here, $Q_{f}({\bf R}_{rs},t;{\bf R}_{rs}^0)$,  the conditional probability for the 
contact distance ${\bf R}_{rs}$ with the initial distribution $P_{I}({\bf R}_{rs}^0)$ of the 
contact distance, ${\bf R}_{rs}^0$, is given in terms of ${\bf C_1}(t)$$=$$\langle {\bf R}_{rs}(t)\rangle$, 
and  $C_2(t)$$=$$\langle[{\bf R}_{rs}(t)-\langle {\bf R}_{rs}(t)\rangle]^2\rangle$ by   
\begin{equation}\label{1.7}
Q_{f}({\bf R}_{rs},t;{\bf R}_{rs}^0)=\left[ {{3}\over {2\pi C_2(t)}}\right]^{{3}\over {2}}
\exp{\left[ -{{3({\bf R}_{rs}-{\bf C_1}(t))^2}
\over {2 C_2(t))}} \right]}
\end{equation}
We define the $N$$\times$$3$-matrices for positions,
${\bf R}(t)$$\equiv$$({\bf R}_1(t),{\bf R}_2(t),\ldots ,{\bf R}_N(t))^\dagger$, random 
forces, ${\bf F}(t)$ $\equiv$$({\bf F}_1(t),{\bf F}_2(t),\ldots ,{\bf F}_N(t))^\dagger$, and the applied force,
${\bf f}$$\equiv$$({\bf 0},{\bf 0},\ldots ,{\bf f}_N)^\dagger$, where ``$\dagger$'' denotes 
the adjoint $3$$\times$$N$ matrix.  The equation of motion in matrix form is
${{d}\over {dt}}{\bf R}+{\bf M}{\bf R}={{1}\over {\xi}}{\bf F}+{{1}\over {\xi}}{\bf f}$
with the $N$$\times$$N$ matrix ${\bf{M}}$$=$$\omega_0({\bf{A}}_0$$+$$\gamma{\bf{A}}_1)$.  The 
tridiagonal Rouse matrix is ${\bf{A}}_0$, with $a^{0}_{1,1}$$=$$a^{0}_{N,N}$$=$$1$, $a^{0}_{m,m}$$=$$2$, 
and $a^{0}_{m,m+1}$$=$$a^{0}_{m-1,m}$$=$$-1$ \cite{20}, and $\omega_0$$=$$3k_B T/2a^2$.  
The interaction between monomers $r$ and $s$ are included in ${\bf{A}}_1$, with 
$a^{1}_{r,r}$$=$$a^{1}_{s,s}$$=$$1$ and $a^{1}_{r,s}$$=$$a^{1}_{r,s}$$=$$-1$. The 
strength of the adhesive contact, relative to the harmonic bond is $\gamma$$=$$(a/b)^2$.

Integration of the equation of motion gives  
${\bf R}(t)$$=$${\bf G}(t)$${\bf R}(0)$$+$$(1/\xi)$$\int_0^tdt'{\bf G}(t')({\bf F}(t')$$+$${\bf f})$, where 
${\bf G}(t)$ is the Greens function; ${\bf G}(\omega)$$\equiv$$\left[ i\omega {\bf I}+{\bf M}\right]^{-1}$ in the Fourier domain. 
To compute ${\bf C_1}(t)$ and $C_2(t)$, we must invert  ${\bf G}(\omega)$. We rewrite 
${\bf A}_1$$=$${\bf P}{\bf P}^{\dagger}$, where $\bf{P}$ is an $N$$\times$$1$ vector with non-zero elements $P_r$$=$$1$ 
and $P_s$$=$$-1$,
and  
${\bf G}(\omega)$$=$${\bf G}_0(\omega)$$-$${\bf G}_0(\omega){\bf P}$$[ {{1}\over {\gamma w_0}}{\bf I}+
{\bf P}^\dagger {\bf G}_0(\omega){\bf P}]^{-1}$${\bf P}^\dagger {\bf G}_0(\omega)$ \cite{242522},
where the matrix elements 
$G_0(\omega)_{mn}$$=$$[i\omega {\bf I}$$+$${\bf M}_0]_{mn}^{-1}$$=$${{1}\over {N}}$$\sum_p{{\cos{[p\pi n/N]}\cos{[p\pi m/N]}}\over {i\omega+\lambda_pw_0}}$ and 
$\lambda_p$$=$$\pi^2 p^2/N^2$ are the  eigenmodes ($p$$=$$0,1,\ldots,N$) \cite{20}.


{\it The pdf of contact lifetimes for a single contact:}
Constraining the first residue of the chain is equivalent to excluding 
$\lambda_0$$=$$0$, which defines the overall translation of the chain. The results are presented by including  
the first ten modes 
for $N$$=$$100$, $a$$=$$0.4nm$, the diffusion coefficient, $D$$=$$10^6$$nm^2/s$, and $k_B T$$=$$4.1$$pN$$nm$. 
We chose $P_{I}({\bf R}_{rs}^0)$$=$${{1}\over {4\pi (B/2)^2}}$$\delta (|{\bf R}_{rs}^0|-B/2)$, 
with the rupture distance set to $B$$=$$2a$.  
To assess the effect of the strength of the adhesive contact 
on the contact rupture kinetics, we used $\gamma$$=$$1$ (strong contact) and $\gamma$$=$$0.01$ (weak contact). 
The critical equilibrium force required to rupture the contact is  $f_c$$=$$(3 k_B T B/b^2)$; $f_c$$=$$31pN$ 
for $\gamma$$=$$1$, and $f_c$$=$$0.31pN$ for $\gamma$$=$$0.01$. The values of $f$ were set to $f$$=$$1pN$ 
(weak force) and $f$$=$$40pN$ (strong force). At $f$$=$$1pN$, $f/f_c$$\ll$$1$ for 
$\gamma$$=$$1$, and $f/f_c$$\sim$$1$ for $\gamma$$=$$0.01$; at $f$$=$$40pN$,  $f/f_c$$>$$1$ for 
$\gamma$$=$$1$ and $f/f_c$$\gg$$1$ for $\gamma$$=$$0.01$.


\begin{figure}[t]
\includegraphics[width=.48\textwidth]{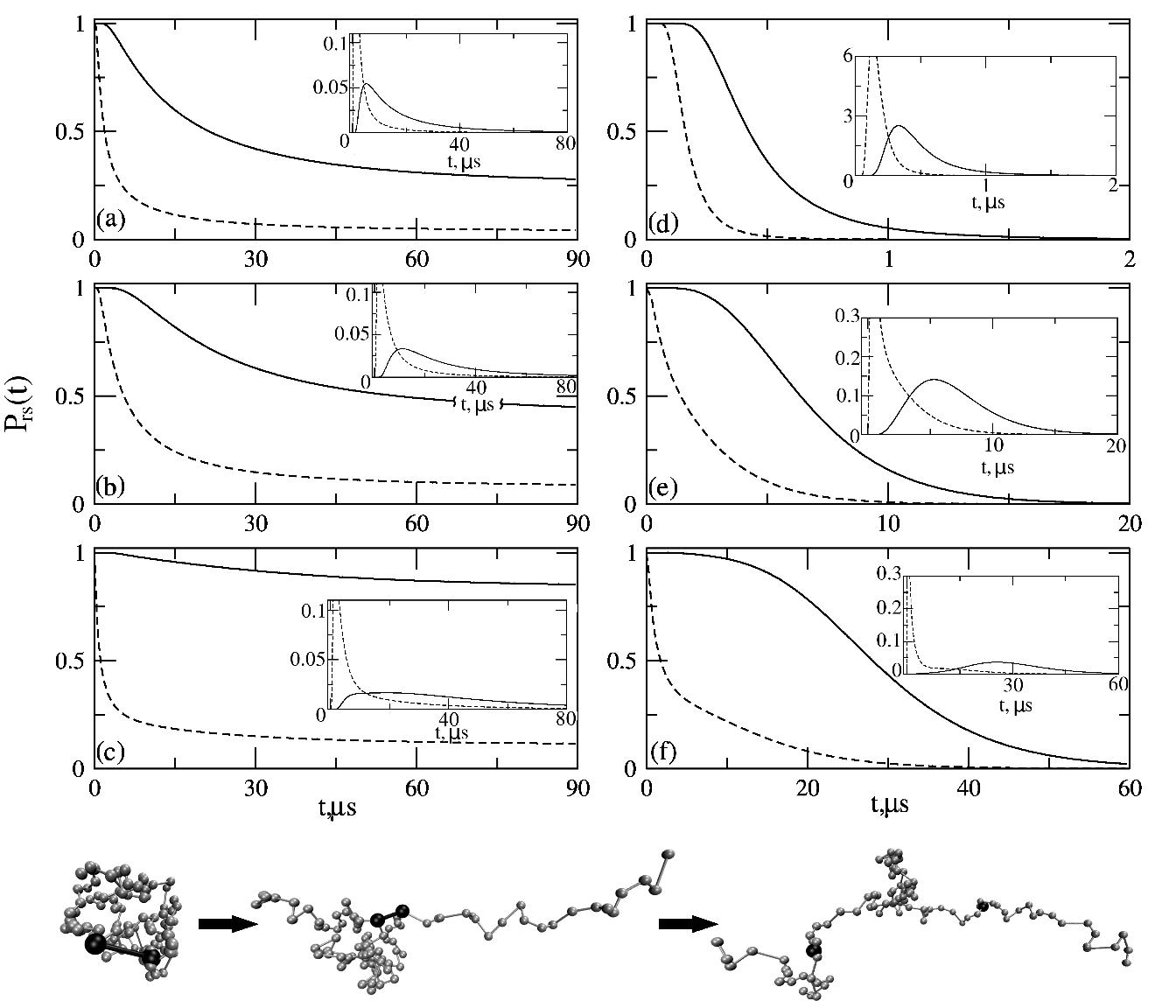}
\caption{The survival probabilities $P_{rs}(t)$ for a single contact $(r,s)$$=$$(1,100)$ 
({{$\mathbf{a}$, $\mathbf {d}$}}), $(20,80)$ ($\mathbf {b}$, $\mathbf {e}$), and $(40,60)$ ($\mathbf {c}$, 
$\mathbf {f}$), with  $ {f}$$=$$1$pN ($\mathbf {a}$$-$$\mathbf {c}$) and $40$pN ($\mathbf d$$-$$\mathbf f$), 
and with $\gamma$$=$$1$ (solid lines) and $0.01$ (dashed lines). The lifetime distributions $p_{rs}(t)$ 
are shown in the insets. Notice the variations in time scales in the panels. The chain structures for the intact single contact $(20,80)$, 
before and after the rupture are shown.}
\end{figure}


The survival probabilities $P_{rs}(t)$ (Fig. 1) of the adhesive contact at $f$$=$$1 pN$ and $f$$=$$40 pN$ 
for $\gamma$$=$$0.01$ and $\gamma$$=$$1$ show striking variations with respect to $|r - s|$.  
For a fixed $f$ (low forces with  $f/f_c$$\ll$$1$ or $\sim$$1$) the decay of $P_{rs}(t)$ not 
only depends on the strength of the contact, but also on its location. At $f$$=$$1pN$, $P_{rs}(t)$ 
for strong and weak bonds display markedly different kinetics, with $P_{rs}(t)$ for weak contact 
$(r,s)$$=$$(1,100)$ and $(20,80)$ decaying to values close to zero on $30 \mu s$ timescale
(Figs. 1a and 1b). However, for $(r,s)$$=$$(40,60)$ the asymptotic value of 
$P_{rs}(t$$\to$$\infty)$$=$$P_{rs}^{eq}$$\approx$$0.12$  (Fig. 1c). Surprisingly, 
$P_{rs}(t)$s for $(r,s)$$=$$(20,80)$ and $(40,60)$ for strong bonds are qualitatively similar (Fig. 1b and 1c) 
and approach higher values of $P_{rs}^{eq}$$\approx$$0.42$ and $0.85$ respectively, whereas 
$P_{rs}^{eq}$$\approx$$0.26$ for $(r,s)$$=$$(1,100)$ (Fig. 1a). Thus, the rupture of contacts 
in the middle of the chain is strongly dependent on the coupling between the internal modes of the chain 
and the dynamics of instability due to the applied force. We refer to this as the 
{\it ``internal motion dominated'' (IMD)} regime. The stochastic nature of the bond rupture kinetics is also reflected 
in the decay of $P_{rs}(t)$ for strong and weak bonds. For $(r,s)$$=$$(1,100)$, $P_{rs}(t)$ decays faster to 
lower $P_{rs}^{eq}$, due to larger chain 
fluctuations at the chain ends compared 
to contacts $(r,s)$$=$$(20,80)$ and $(40,60)$ in the middle of the chain. In contrast to low 
forces, for $f/f_c$$>$$1$,  $P_{rs}(t)$ for all contacts decay to zero at long times, regardless of the  contact strength 
(Figs. 1d-1f). However, $P_{rs}(t)$ for strong contact $(r,s)$$=$$(40,60)$, which 
is close to the midpoint of the chain, decays to zero on a much slower $40\mu s$ timescale, compared 
to $10\mu s$ and $1\mu s$ timescale for $(r,s)$$=$$(20,80)$ and $(1,100)$. At $f$$=$$40pN$, 
$P_{rs}(t)$ for weak contacts decay to zero on similar time scales (Figs. 1d-1f) which shows that at large 
forces rupture kinetics is solely determined by the instability caused by the applied tension, with chain 
dynamics playing relatively minor role. Thus, as $f/f_c$ increases, the rupture kinetics become increasingly 
{\it ``tension dominated''}, i.e. contact rupture is due to the applied force. Interestingly, there is a significant 
plateau in the decay of $P_{rs}(t)$ especially for $\gamma$$=$$1$. The duration of the plateau increases 
as $|r-s|$ decreases and $f$ increases (Figs. 1d-1f), and shows  that the kinetics of rupture is non-exponential. 
  
The lifetime distributions, $p_{rs}(t)$ (insets in Fig. 1), show longer tails for the stronger contacts;   
$p_{rs}(t)$ for the interior contact $(40,60)$ (Figs. 1c and 1f), does not have a pronounced  dependence 
on $f$ compared to $p_{rs}(t)$ for contacts $(20,80)$ and $(1,100)$ (Figs. 1a, 1b, 
and 1d and 1e). In particular,  $p_{rs}(t)$ at $f$$=$$1pN$ and $f$$=$$40pN$ \textit{agree quantitatively} 
for  strong and weak $(40,60)$-contacts. This implies that the disruption of the interior contacts is 
mediated by internal chain motions, and hence is in the IMD regime even at 
high $f$. The influence of internal motion of the chain on the contact rupture kinetics is also reflected in the 
width of $p_{rs}(t)$s which  broadens for remote contacts $(20,80)$ 
and $(40,60)$.


\begin{figure}
\includegraphics[width=.48\textwidth]{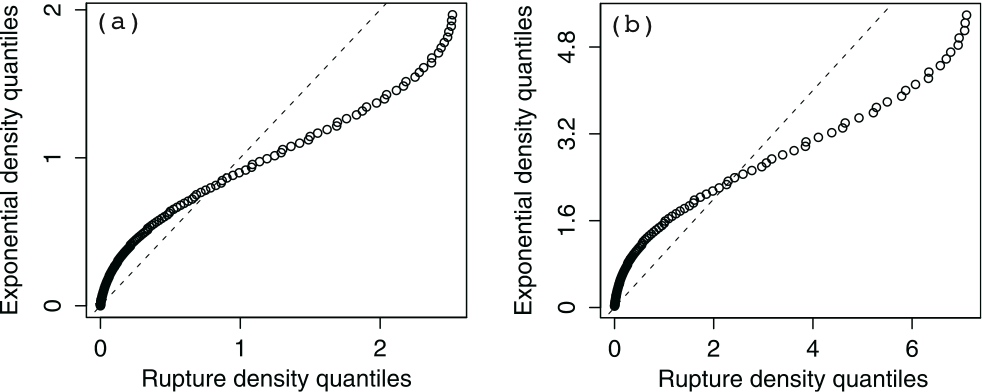}
\caption{The $Q$$-$$Q$ plots (empty circles) of lifetimes $p_{rs}(t)$ for single contact $(r,s)$$=$$(1,100)$ 
($x$-axis) versus  lifetimes  $p_{exp}(t)$ ($y$-axis), for $f$$=$$40$pN and $\gamma$$=$$1$ ($\mathbf a$) 
and $\gamma$$=$$0.01$ ($\mathbf b$). The rate $K$ in $p_{exp}(t)$ is set to the inverse of the average 
contact lifetime (the maximum likelihood estimate) \cite{1919a}.}
\end{figure}


{\it Contact lifetime pdf is non-exponential:}
Forced rupture of noncovalent bonds is often described using two-state kinetics, $(rs)$$\to$$0$, which 
corresponds to the exponential pdf of contact lifetimes, $p_{exp}(t)$$=$$K$$\exp{[-Kt]}$ where $K$ is 
rupture rate.  The results for $p_{rs}(t)$ in Fig. 1 deviate from $p_{exp}(t)$, especially when $f/f_c$$<$$1$ 
or when the contact $(rs)$ is in the chain interior.  The extent of deviation is assessed using the 
quantile-quantile ($Q$$-$$Q$) plot \cite{1919a,Barsegov07BiophysJ}. A quantile is a number $x_p$ such that 
$100$$\times$$p\%$ of the probability values are $\le$$x_p$. The $Q$$-$$Q$ scatterplot 
compares two sets of probability values: ${x_p}$, the quantile for $p_{rs}(t)$, and ${y_p}$, the quantile for $p_{exp}(t)$.
If the two sets are similarly distributed, the points fall on the reference line \cite{1919a}. The 
$Q$$-$$Q$ plots of the lifetime distribution for the contact $(r,s)$$=$$(1,100)$ at $f$$=$$40pN$ and for 
$\gamma$$=$$1$ and $\gamma$$=$$0.01$ show that $p_{rs}(t)$ is larger (smaller) than $p_{exp}(t)$ in 
the middle range (tail) of the lifetimes (Fig. 2). The non-exponentiality arises because  it takes time 
($\tau_{tp}$) for the force-induced tension to propagate to the contact \cite{Structure}. 

{\it Interplay of timescales:} 
The  non-exponential nature of $p_{rs}(t)$ is determined by the interplay of two 
timescales,  $\tau_{tp}$ and the spectrum of relaxation times $\{\tau_p\}$ 
($p$$=$$1,2,\ldots, N$). These determine the evolution of
${\bf C}_1(t)$ due to the applied force ${\bf f}$, and the broadening of the Gaussian 
density, $C_2(t)$, due to the random force ${\bf F}$ (Eq. (\ref{1.7})). To reveal the richness in the rupture 
kinetics, we consider $P_{rs}(t)$ for the $(r,s)$-contacts close to the 
chain ends in the limits  $f/f_c$$<$$1$, and  $f/f_c$$>$$1$. For these contacts, 
the contribution from even Rouse modes is negligible \cite{20}, and only the slowest mode with relaxation 
time $\tau_1$$=$$\xi/\lambda_1$ contributes significantly. As a result, 
$P_{rs}(t)$$=$$c_0\int_{0}^{B}d{\bf R}_{rs}$$\exp{\left[ - {{3({\bf R}_{rs}-{\bf C_1}(t))^2} \over {2 C_2(t))}}\right]}$,
where $c_0$ is the constant, ${\bf C}_1(t)$$\approx$${\bf R}_{rs}^{eq}$$(1-e^{-t/\tau_1})$, 
$C_2(t)$$\approx$$(\Delta{\bf R}_{rs}^{eq})^2$$(1-e^{-2t/\tau_1})$, and ${\bf R}_{rs}^{eq}$ and 
$\Delta{\bf R}_{rs}^{eq}$ denote the equilibrium $(r,s)$-contact vector and fluctuations, respectively.

{\it 1. Weak force, $f/f_c$$<$$1$:} In this regime,  $|{\bf C}_1(t)|$$<$$B$, and hence the kinetics of 
$P_{rs}(t)$ is dominated by the broadening of $C_2(t)$. By evaluating $P_{rs}(t)$ and using the series
expansion of the error function, 
we obtain $P_{rs}(t)$$\sim$$\exp{[-B^2/2C_2(t)]}$. At short times, $t$$\ll$$\tau_1$, and we can Taylor 
expand the exponential function in $C_2(t)$, which to the first order in $t$$\ll$$\tau_1$ reads 
$e^{-2t/\tau_1}$$\approx$$1$$-$$2t/\tau_1$. Then, $P_{rs}(t)$$\sim$$e^{-c_1\tau_1/t}$ ($c_1$ 
is a constant). At longer times, when $t$$\gg$$\tau_1$, $e^{-2t/\tau_1}$$\ll$$1$, and 
$P_{rs}(t)$$\sim$$\exp{[-c_2e^{-2t/\tau_1}]}$ ($c_2$ is a constant). This implies that for a weak 
force, in the long time limit, $P_{rs}(t)$ of the contacts close to the chain 
ends is described by the double-exponential (Gumbel type) distribution of the largest value for the 
exponential density \cite{Gumbel}. 

{\it 2. Strong force, $f/f_c$$>$$1$:} The kinetics of $P_{rs}(t)$ is dominated by the
the dynamics of ${\bf C}_1(t)$.  However, the decay of 
$P_{rs}(t)$ is delayed by $\tau_{tp}$.  The tension propagation timescale $\tau_{tp}$$\approx$$\tau_1$. 
{\it a. Tension propagation regime:} ($t$$<$$\tau_{tp}$). Since $|{\bf C}_1(t)|$$<$$B$, the contact lifetime 
is strongly linked to the random motions of the polymer, which results in broadening of the Gaussian 
density due to increase in $C_2(t)$.  Evaluating $P_{rs}(t)$,  and assuming 
$|{\bf C}_1(t)|$$\approx$$B$$<$$C_2(t)$, 
we obtain $P_{rs}(t)$$\sim$$(t/\tau_1)^3$$e^{-c_3(t/\tau_1)^{1/2}}$ ($c_3$ is a constant). Thus,
 $P_{rs}(t)$ scales with force as $P_{rs}(t)$$\sim$$f^3$$e^{-c_4 f}$ ($c_4$ is a constant). 
{\it b. Kinetic regime: ($t$$>$$\tau_{tp}$)}. The contact lifetime is 
still linked to the random motions of the polymer (broadening of $C_2(t)$) but is dominated by
 the dynamics of ${\bf C}_1(t)$$>$$B$. 
In this regime, $e^{-t/\tau_1}$$\ll$$1$ in ${\bf C}_1(t)$ and $e^{-2t/\tau_1}$$\ll$$1$ in $C_2(t)$, 
and $P_{rs}(t)$ is again double-exponential (Gumbel) density \cite{Gumbel}, i.e. 
$P_{rs}(t)$$\sim$$\exp{[-c_5 e^{-2t/\tau_1}]}$, where $c_5$ is a constant. We find
$P_{rs}(t)$$\sim$$f^2$$e^{-c_6 f}$, where $c_6$ is a constant.


{\it The contact lifetimes for two contacts:}
The Hamiltonian of the chain with two contacts $(r,s)$ and $(r',s')$ is given by
$H$$=$${{3k_B T}\over {2a^2}}$$\sum_{n=2}^{N}{\bf R}_{n,n-1}^2$$+$${\bf f}$${\bf R}_{N1}$$+$
$\sum_{(i,j)}$${{3k_B T}\over {2b_{ij}^2}}$$\Theta(B_{ij}$$-$$|{\bf R}_{ij}|){\bf R}_{ij}^2$,
where $b_{ij}$ ($B_{ij}$) are the equilibrium (critical rupture) distance, 
which we rewrite  as 
$H$$=$$H_{rs}$$+$$H_{r's'}$, where 
$H_{ij}$$=$${{1}\over {2}}$${{3k_B T}\over {2a^2}}$$\sum$${\bf R}_{n,n-1}^2$$+$${{{\bf f}}\over {2}}$${\bf R}_{N1}$$+$${{3k_B T}\over {2b_{ij}^2}}$$\Theta(B_{ij}$$-$$|{\bf R}_{ij}|)$${\bf R}_{ij}^2$. 
The population is
$P(t)$$=$$P_{rs,r's'}(t)$$+$$P_{0,r's'}(t)$$+$$P_{rs,0}(t)$, where
$P_{rs,r's'}(t)$$=$$P_{rs}(t)$$P_{r's'}(t)$ denotes that both contacts are intact,  
$P_{0,r's'}(t)$$=$$P_{rs\to 0}(t)$$P_{r's'}(t)$ means that $(r,s)$ is ruptured but $(r',s')$ 
is intact, and  $P_{rs,0}(t)$$=$$P_{rs}(t)$$P_{r's'\to 0}(t)$ implies that $(r',s')$ is disrupted 
but $(r,s)$ is intact. $P_{ij \to 0}(t)$$=$$1$$-$$P_{ij}(t)$ is the population 
of conformations with the disrupted contact $(i,j)$. 
Using  $r'$$\approx$$r$, $s'$$\approx$$s$, and $b_{rs}$$\approx$$b_{r's'}$$=$$b$, the Hamiltonian 
for two nearest neighbor contacts is 
$H$$=$${{3k_B T}\over {2a^2}}$$\sum$${\bf R}_{n,n-1}^2$$+$${\bf f}$${\bf R}_{N1}$
$+$${{3k_B T}\over {2\bar{b}^2}}$$\theta(B$$-$$|{\bf R}_{rs}|)$${\bf R}_{rs}^2$, 
where 
$\bar{b}$$=$$b/\sqrt{2}$ is the rescaled distance. $P_{tot}(t)$ ($p_{tot}(t)$) 
is computed using $P_{rs}(t)$ ($p_{rs}(t)$) for single contact $(r,s)$.


\begin{figure}[t]
\includegraphics[width=.48\textwidth]{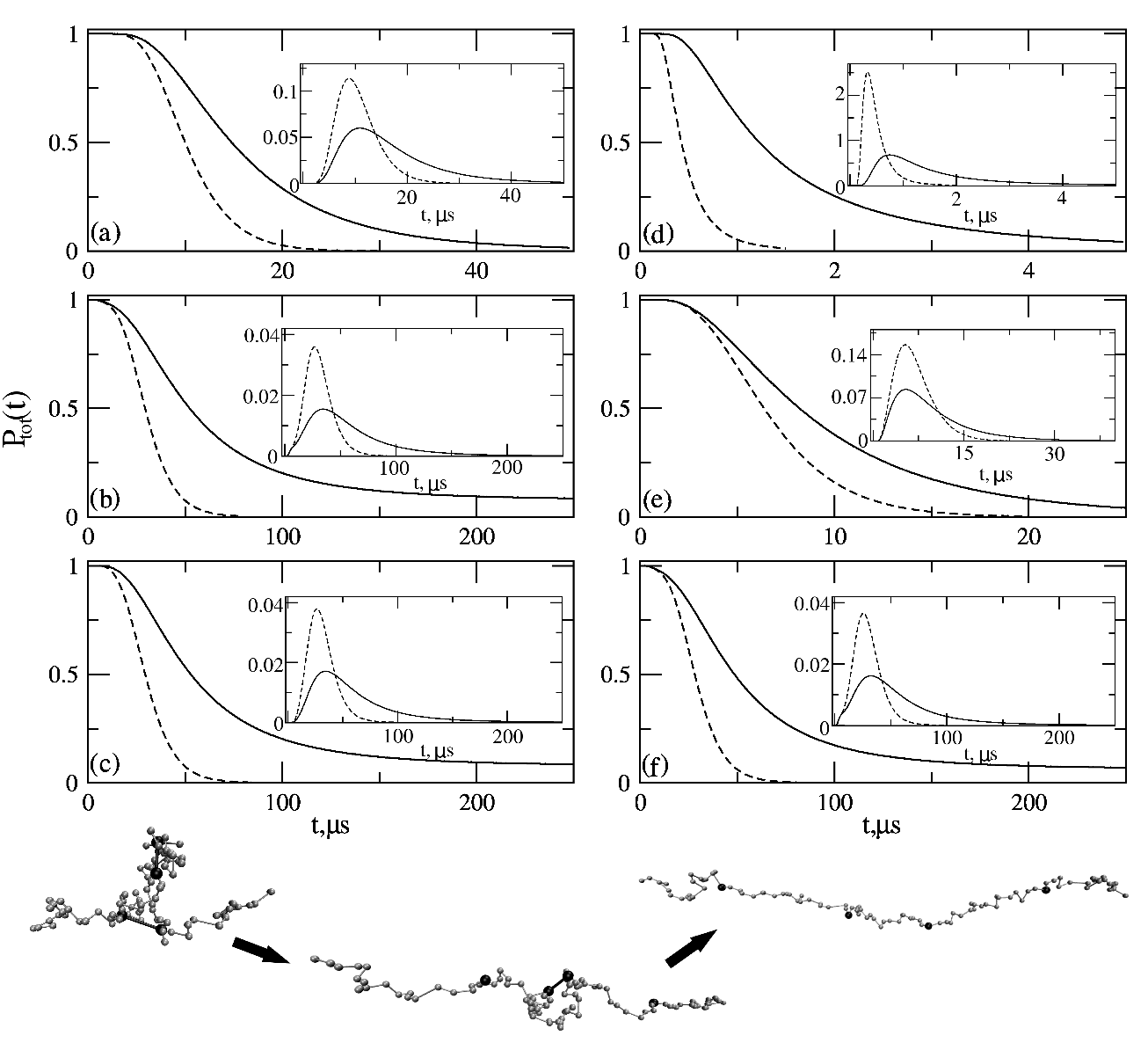}
\caption{The survival probabilities $P_{tot}(t)$ for  contact pairs $(1,100)$ and $(20,80)$
($\bf{a}$), $(1,100)$ and $(40,60)$ ($\bf{b}$), and $(20,80)$ and $(40,60)$ ($\bf{c}$), 
$(1,100)$ and $(2,99)$ ($\bf{d}$), $(20,80)$ and $(21,79)$ ($\bf{e}$), and $(40,60)$ 
and $(41,59)$ ($\bf{f}$), for $f$$=$$40$pN (solid lines) and $80$pN (dashed lines), all with $\gamma$$=$$1$. 
The lifetime distributions $p_{tot}(t)$ are shown in the insets.  Time scales vary greatly in all panels. Structures are for the intact pair $(20,80)$
$(40,60)$, disrupted contact  $(20,80)$ and intact contact $(40,60)$, and both contacts disrupted are shown.}
\end{figure}


Plots of $P_{tot}(t)$ for two separated strong contacts at $f$$=$$40pN$ and $80pN$ 
show that, in general, increasing $f$ facilitates more rapid rupture of the binary contacts (Figs. 3a-3c). 
The presence of interior contacts  delays their rupture 
(compare Figs. 3a with 3b and 3c) even though $f/f_c$$>$$1$.  
The time dependence of $P_{tot}(t)$ for two separated contacts (Figs 3a-c), after a lag phase, can be 
analyzed using two distinct exponential functions, which shows the rupture occurs by sequential kinetics, 
with  $\{(r,s),(r',s']\}$$\to$$\{(0),(r',s')\}$$\to$$\{(0),(0)]\}$.
The {\it rate limiting second step} is the rupture of the more interior contact. Comparison of  $P_{tot}(t)$, 
for $\gamma$$=$$1$ and $f$$=$$40pN$, for the nearest neighbor contact pairs $(r,s)$$=$$(1,100)$ and 
$(r',s')$$=$$(2,99)$, $(20,80)$ and $(21,79)$, and $(40,60)$ and $(41,59)$ (Fig. 3d-3f) with $P_{rs}(t)$ 
for single contact $(r,s)$$=$$(1,100)$, $(20,80)$, and $(40,60)$ 
(Fig. 1d-1f) shows that the double bond increases stability of the contact, especially for the
interior contact $(40,60)$.   The chain with double bonds ruptures in a single step, 
$\{(r,s),(r',s')\}$$\to$$\{(0),(0)\}$, i.e. the rip occurs cooperatively, so that $P_{tot}(t)$ decays with
 a single rate constant at long times. 
The decay of $P_{tot}(t)$ for separated contacts, 
$(r,s)$$=$$(20,80)$ and $(r',s')$$=$$(40,60)$, is slower than $P_{tot}(t)$ for the nearest neighbor 
contacts, $(1,100)$ and $(2,99)$ (Figs. 3c and 3d), and $P_{tot}^{eq}$ is larger, which 
shows that the persistence of binary interactions critically depends on the location of contacts 
in the chain. 
The lifetime distributions show  that increasing the force from $f$$=$$40pN$ to $f$$=$$80pN$ results 
in the decreased stability of contacts and shorter lifetimes (Fig. 3). 
The contact lifetimes increase when they are located in the chain interior (Fig. 3).
The lifetimes of the nearest neighbor contacts, $(1,100)$ and $(2,99)$ (Figs. 3d and 3e),  are 
shorter than that for the pair, $(40,60)$ and $(41,59)$ (Fig. 3f), which shows
the importance of the location of binary contacts with respect to the point of application of the force. 


{\it Conclusions:}    The present work shows that the shape of the free energy landscape  can only be 
discerned by analyzing the topology-dependent lifetime distributions of adhesive contacts that reflect 
the structure the molecules.  The continued increase in the temporal resolution in laser optical
tweezer experiments should allow the prediction that, the internal motions of biomolecules are coupled to global fluctuations, to be tested. 
The present theory provides a conceptual framework for interpreting such experiments \cite{BlockAnnRevBiophys2007TinocoPNAS07FernandezPNAS07}. In contrast, the rich  behavior predicted here cannot
be obtained by solving for the dynamics in the exactly calculable equilibrium free energy profile in the presence of non-zero force. Accounting for the non-exponential distribution of unbinding lifetimes 
will require models that reflect the interplay between local chain motions and tension-induced global 
unfolding.

{\bf Acknowledgments:} This work was supported by National Science Foundation Grant NSFCHE-05-14056.

\end{document}